# Change-point analysis in frequency domain for chronological data

Gyorgy H. Terdik

*Faculty of Informatics, University of Debrecen*

*Debrecen, Hungary*

Stergios B. Fotopoulos*

*Department of Finance and Management Science, Carson College of Business*

*Washington State University*

Venkata K. Jandhyala

*Department of Mathematics and Statistics, College of Arts & Sciences*

*Washington State University*

November 18, 2016

**Summary**

The purpose of this study is to provide a new methodology of how one can consistently estimate a change-point in time series data. In contrast with previous studies, the suggested methodology employs only the empirical spectral density and its first moment. This is accomplished when both the means and variances before and after the unidentified time point are unknown. Then, the well-known Gauss-Newton algorithm is applied to estimate and provide asymptotic results for the parameters involved. Simulations carried out under different distributions, sizes and unknown time points confirm the validity and accuracy of the methodology. The real-world example considered in the paper illustrates the robustness of the methodology in the presence of even extreme outliers.

## 1. Introduction

Spectral analysis and frequency domain methods play a central role in nonparametric analysis of time series data. The function applied to analyze the change-point in a time series is based on the spectral density. The relevance of empirical spectral processes to both stationary and non-stationary time series has been growing rapidly in the last couple of decades. The role of the empirical spectral density for independently distributed data with means different from zero is the theme of this article. Specifically, the goal here is to provide an alternative and robust methodology for consistently estimating an unknown time point in a time series such that both the mean and variance have been significantly altered from past to present interval sets of data. The method where a change-point is estimated is based on the exploitation of the frequency domain approach. Detecting an unknown time point in a time series under frequency domain when the characteristics of the process have changed has been considered by Picard (1985) and Giraitis and Leipus (1990, 1992). Recently, empirical spectral processes or frequency domain practice were introduced for distinguishing short-range dependence with a single change point (CP), and long-range dependence (LRD) in time series. The confusion between LRD and non-stationary alternatives has been well documented in the literature. In particular, Shimotsu (2006), Ohanissian et al. (2008), Müller and Watson (2008) and Qu (2010) assume LRD as the null hypothesis versus the alternative being a non-stationary time series. Tests where a non-stationary model is null have also been considered by Berkes et al. (2006) Jach and Kokoszka (2008) and Yau and Davis (2011).

In this paper we consider the CP model only, and we are not interested in answering how to resolve the ambiguity between LRD and CP models. Although it is reasonable to assert that an element of vagueness exists between these two models, i.e., share similar properties especially within the spectrum, this paper addresses how one can consistently estimate the CP parameters and determine their asymptotic distribution. It must be understood that the data derived from either one of these two models could lead to a misspecification. The assumption here is that the data is filtered sufficiently and shows and is strongly in favor of the CP model.

The problem of change-point detection and estimation has been actively studied over the last several decades in statistics. A typical statistical formulation of change-point detection is to consider probability distributions from which data in the past and present intervals are generated and regards the target time point as a change point if the two distributions are significantly different. Various approaches to change-point detection and estimation have been investigated within this statistical framework, including the CUSUM and GLR approaches, maximum likelihood estimation (mle), nonparametric, semiparametric, just to name a few. In particular, maximum likelihood estimation of the change-point under the time domain formulation has been first studied by Hinkley (1970, 1972). Recent advances for the asymptotic distribution of the change-point mle have been considered by Jandhyala and Fotopoulos (1999), Borovkov (1999), Jandhyala and Fotopoulos (2001), Fotopoulos and Jandhyala (2001), Fotopoulos (2009) and Fotopoulos et al (2010). In recent years, there have been several publications on spectrum-based methods for change-point detection where the changes in autocorrelation structure are caught through the Fourier or wavelet-based spectrum analysis, Adak (1998), Ombao et al. (2001), Choi and Ombao (2008). As is often the case, a data set has a known explicit model from which it is generated; analysts then look to fit an appropriate model or models to such a series in the hope of understanding the underlying mechanism. In a series of chronological observations being independently distributed, we provide a new estimation methodology when the characteristics in the past and present intervals have been significantly altered. The method is based on utilizing the deviation of the empirical from the spectral density. It is shown here that under mild moment conditions the estimates of the parameters are weakly consistent and follow asymptotic normality. The technology employed to establish these results is 1) the use of the first moment of the periodogram and 2) the Gauss-Newton algorithm.

The organization of this study is as follows. Section 2 develops properties related to the change-point problem using the periodogram. The first two moments of the periodogram are computed under the presence of an unknown parameter change point. Section 3 provides key results of how one can consistently obtain estimates of the change-point and other key parameters. The robustness of the methodology is attained using numerous simulations under various distributions in Section 4. Finally, Section 5 provides an explicit example which shows that the methodology is robust even under the presence of extreme outliers.

## 2. Problem formulation and basic approach

Let $\mathscr{G}$ denote a set of real-valued stationary processes with absolutely continuous spectral measure, which satisfies at least four moment conditions. Let $f_X$ be its spectral density with respect to the Lebesgue measure $d_\varsigma(\lambda) = d\lambda$ on the torus $[-\pi, \pi]$.

The spectrum densities before and after the change-point time will be denoted by $f_B(\cdot)$ and $f_A(\cdot)$, respectively (the letter B will indicate before the change-point time and A after in future notations). The same will apply for the vector of observations, means variances and any other characteristic involved in this formulation. Note that the time-point is unknown.

Consider $\{X_t : t = 0, 1, \cdots, T-1\}$ as a restriction to $\{0, 1, \cdots, T-1\}$ of a stationary time series belonging to $\mathscr{G}$. We formulate the following hypothesis.

$H_0$: $X_t : t = 0, 1, \cdots, T-1$ represents observations belonging to $\mathscr{G}$, against

$H_A$: there exists a time-point $\tau \in \{1, \cdots, T-2\}$, $0 < \tau < T-1$, such that $X_j = X_j^{(B)}$, $j \leq \tau - 1$, with $E[X_j^{(B)}] = \mu_B$, and spectrum $f_B(\cdot)$ and $X_j = X_j^{(A)}$ if $j > \tau - 1$, with $E[X_j^{(A)}] = \mu_A$, and spectrum $f_A(\cdot)$, $\mu_B \neq \mu_A$ and $f_B(\omega) \neq f_A(\omega)$, for some $\omega \in [-\pi, \pi]$.

Here, both $X_j^{(B)}$ and $X_j^{(A)}$ are in $\mathscr{G}$, with spectral density measures $f_B(\cdot)$, $f_A(\cdot)$, respectively.

Assume that there exists $\lambda \in (0,1)$ such that $\tau = \lfloor \lambda T \rfloor$ with $\lfloor x \rfloor$ denoting the integer part of $x$. Write $\lambda_T = \tau/T \to \lambda$ as $T \to \infty$, i.e., $\lambda$ is the relative location of the change-point model. In other words, as the sample size $T$ increases, $\lambda$ remains fixed so that the proportion of the lengths of both segments remains the same. Throughout this section, $\lambda$ is treated as a parameter. The problem of estimating $\lambda$, as well as other process parameters under the spectral domain, is the main contribution of this article and it will be considered in the subsequent section. For asymptotic purposes, we let the lengths of the two segments $\tau$ and $T-\tau$ to go to infinity at the same rate as $T$, the total number of observations ($\tau = O(T)$ and $T - \tau = O(T)$). As usual, let $\langle \cdot, \cdot \rangle$ denote the inner product.

We begin by considering the discrete Fourier transform defined by

$$d_X^{(T)}(\omega) = \langle \underline{X}_T, \underline{e}_T \rangle = \sum_{t=0}^{T-1} X_t e^{it\omega_k^{(T)}}, \tag{2.1}$$

where $\underline{X}'_T = (X_0, X_1, \cdots, X_{T-1})$, $\underline{e}'_d(\omega_k^{(d)}) = (1, e^{i\omega_k^{(d)}}, \cdots, e^{i(d-1)\omega_k^{(d)}})$, and Fourier frequencies $\omega_k^{(d)} = 2\pi k/d$ for a given sample size $d$ and $k = 0, 1, \cdots, d-1$. Note that the vector $\underline{e}_d(\omega_k^{(d)})$ constitutes an orthonormal basis for the $d$-dimensional complex space $\mathbf{C}^d$.

Under the change-point $\tau$, the process $X_t$, $t = 0, 1, \cdots, T-1$, is not stationary. Nevertheless, one can represent $X_t$ as a linear combination of two stationary processes through a taper. Thus, we let $h$ be a taper defined on $[0,1]$ as $h(s) = 1$, for $s \in [0, \lambda_T]$, and $h(s) = 0$, for $s \in (\lambda_T, 1]$. Therefore, the discrete process $X_t$ is represented in terms of the taper $h$ as

$$X_t = h(t/T) X_t^{(B)} + \{1 - h(t/T)\} X_t^{(A)}. \tag{2.2}$$

Let the vectors $\underline{X}_T$ and $\underline{e}_T(\omega_k^{(T)})$ be partitioned as

$$\underline{X}'_T = \left( \underline{X}^{(B)'}, \underline{X}^{(A)'} \right) \text{ and } \underline{e}'_T(\omega_k^{(T)}) = \left( \underline{e}'_\tau(\omega_k^{(T)}), e^{i\tau\omega_k^{(T)}} \underline{e}'_{T-\tau}(\omega_k^{(T)}) \right),$$

where $\underline{e}'_\tau(\omega_k^{(T)}) = (1, e^{i\omega_k^{(T)}}, \cdots, e^{i(\tau-1)\omega_k^{(T)}})$, and $\underline{e}'_{T-\tau}(\omega_k^{(T)}) = (e^{i\omega_k^{(T)}} \cdots, e^{i(T-\tau)\omega_k^{(T)}})$. Hence, the discrete Fourier transform in (2.1) is analogously partitioned as

$$d_{\underline{X}}^{(T)}(\omega_k^{(T)}) = \langle \underline{X}_T, \underline{e}_T(\omega_k^{(T)}) \rangle = \left\langle \left( \underline{X}^{(B)'}, \underline{X}^{(A)'} \right), \left( \underline{e}'_\tau(\omega_k^{(T)}), e^{i\tau\omega_k^{(T)}} \underline{e}'_{T-\tau}(\omega_k^{(T)}) \right) \right\rangle$$

$$= \langle \underline{X}^{(B)}, \underline{e}_\tau(\omega_k^{(T)}) \rangle + e^{i\tau\omega_k^{(T)}} \langle \underline{X}^{(A)}, \underline{e}_{T-\tau}(\omega_k^{(T)}) \rangle = d_{\underline{X}^{(B)}}^{(T)}(\omega_k^{(T)}) + e^{i\tau\omega_k^{(T)}} d_{\underline{X}^{(A)}}^{(T)}(\omega_k^{(T)}). \tag{2.3}$$

From Theorem 4.4.2 in Brillinger (1981), the process $d_{\underline{X}}^{(T)}(\omega_k^{(T)})$ is asymptotically complex Gaussian. When $2\omega_j \neq 0 \mod(2\pi)$ and $\omega_j \pm \omega_k \neq 0 \mod(2\pi)$, one has,

$$\lim_{T \to \infty} T^{-1} \text{Cov}\left( d_{\underline{X}^{(x)}}^{(T)}(\omega_j), d_{\underline{X}^{(x)}}^{(T)}(\omega_k) \right) = 0, \quad x = B, A. \tag{2.4}$$

From (2.4), it can be concluded that $d_{\underline{X}^{(x)}}^{(T)}(\omega_j)$ and $d_{\underline{X}^{(x)}}^{(T)}(\omega_k)$ are asymptotically orthogonal. In addition, when $\omega_j \neq 0 \mod(2\pi)$, the asymptotic variance is given by

$$\lim_{T \to \infty} T^{-1} \text{Cov}\left( d_{\underline{X}^{(B)}}^{(T)}(\omega_j), d_{\underline{X}^{(A)}}^{(T)}(\omega_j) \right) = 2\pi H_{B,A}(0) f_{B,A}(\omega_j), \tag{2.5}$$

where $f_{B,A}$ denotes the cross spectral density measure between $\underline{X}^{(B)}$ and $\underline{X}^{(A)}$. The expression $H_{B,A}$ is defined as (see, e.g., Brillinger, 1981) $H_{B,A}(\omega) = \int_0^1 h(t)\{1-h(t)\}e^{-it\omega}dt$. In particular, $H_{B,A}(0) = 0$, and

$$H_B = H_{B,B}(0) = \lim_{\omega \to 0} \int_0^1 h^2(t)e^{-it\omega}dt = \lambda, \quad H_A = H_{A,A}(0) = \lim_{\omega \to 0} \int_0^1 \{1-h^2(t)\}e^{-it\omega}dt = 1-\lambda.$$

In light of the above preliminary results, the asymptotic variance-covariance matrix of $d_{\underline{X}}^{(T)}(\omega)$ under the change-point for fixed frequency $\omega \in [0, 2\pi]$ is diagonal. This leads to the following proposition.

**Proposition 2.1.** *Let $\lambda_T \in (0,1)$ and let $\underline{X}_B = (X_0, X_1, \cdots, X_{\tau-1})$ and $\underline{X}_A = (X_\tau, X_{\tau+1}, \cdots, X_{T-1})$ be in $\mathscr{G}$. The two successive stationary sequences are chosen such that the expected values and the spectral densities are $E[X_i] = \mu_B$, and spectrum $f_B(\omega_j)$, for $i \leq \tau - 1$, and $E[X_i] = \mu_A$, and spectrum $f_A(\omega_j)$, for $i \geq \tau$, respectively. Then the Fourier transform*

$$d_{\underline{X}}^{(T)}(\omega_k^{(T)}) = d_{\underline{X}^{(B)}}^{(T)}(\omega_k^{(T)}) + e^{i\tau\omega_k^{(T)}} d_{\underline{X}^{(A)}}^{(T)}(\omega_k^{(T)})$$

*is asymptotically complex Gaussian. Also, if $\lim_{T \to \infty} \lambda_T = \lambda$, the following statements hold:*

1. *For $\omega_j, \omega_k \in [0, 2\pi]$, with $2\omega_j, \omega_j \pm \omega_k \neq 0 \mod(2\pi)$, the following holds:*

$$\lim_{T \to \infty} T^{-1} \text{Cov}(d_{\underline{X}}^{(T)}(\omega_j), d_{\underline{X}}^{(T)}(\omega_k)) = 0$$

2. *For $\omega_j \in [0, 2\pi]$, with $\omega_j \neq 0 \mod(2\pi)$, the following holds:*

$$\lim_{T \to \infty} T^{-1} \text{Cov}(d_{\underline{X}}^{(T)}(\omega_j), d_{\underline{X}}^{(T)}(\omega_j)) = 2\pi\lambda f_B(\omega_j) + 2\pi(1-\lambda) f_A(\omega_j).$$

2.2. *The periodogram and their moment properties.* Write, as above, $H_s(\omega_j)$, for $s = B, A$. The following lemma is of importance.

**Lemma 2.1.** *Under h being the indicator variable, the following holds:*

1. *For $\omega = 0 \mod(2\pi)$, $|d_{\underline{X}}^{(\tau)}(\omega)| = \tau$.*

2. *For fixed $\omega \neq 0 \mod(2\pi)$, $|d_{\underline{X}}^{(\tau)}(\omega)| \leq 1/|\sin(\omega/2)|$.*

**Proof.** Let $\omega \neq 0 \mod(2\pi)$ be fixed. Then,

$$\left|d_{\underline{X}}^{(\tau)}(\omega)\right| = \left|\frac{1-e^{-i\tau\omega}}{1-e^{-i\omega}}\right| \leq 1/|\sin(\omega/2)|,$$

which is bounded for fixed $\omega \in (0, 2\pi)$. □

To develop asymptotic moment properties under a change point for the periodogram of the vector $\underline{X}_T$, at each discrete frequency $\omega_k^{(T)} = 2\pi k/T$, $k = 0, 1, \cdots, d-1$, let $I_{\underline{X}}^{(T)}(\omega_k^{(T)})$ denote the periodogram which is represented in terms of the discrete Fourier transform by

$$I_{\underline{X}}^{(T)}(\omega_k^{(T)}) = T^{-1} \left| \langle \underline{X}_T, e_T(\omega_k^{(T)}) \rangle \right|^2 = T^{-1} \left| \sum_{t=1}^{T-1} X_t e^{it\omega_k^{(T)}} \right|^2 = T^{-1} \left| d_{\underline{X}}^{(T)}(\omega_k^{(T)}) \right|^2. \quad (2.6)$$

The following theorem holds.

**Theorem 2.1.** *Let* $\lambda_T \in (0,1)$ *and let* $\underline{X}_B = (X_0, X_1, \cdots, X_{\tau-1})$ *and* $\underline{X}_A = (X_\tau, X_{\tau+1}, \cdots, X_{T-1})$ *be in* $\mathscr{G}$.
*Then, the expected periodogram is given by*

$$E[I_{\underline{X}}^{(T)}(\omega_k^{(T)})] = 2\lambda_T \pi f_B(\omega_k^{(T)}) + 2\pi(1-\lambda_T) f_A(\omega_k^{(T)}) + |\mu_B - \mu_A|^2 \frac{\sin^2(\tau \omega_k^{(T)}/2)}{T \sin^2(\omega_k^{(T)}/2)} + o(1),$$

*where* $k \neq 0 \mod(T)$.

**Proof.** From (2.6) it follows that

$$TE[I_{\underline{X}}^{(T)}(\omega_k^{(T)})] = E\left[ \left| d_{\underline{X}}^{(T)}(\omega_k^{(T)}) \right|^2 \right] = \mathrm{Cov}\left( d_{\underline{X}}^{(T)}(\omega_k^{(T)}), \overline{d_{\underline{X}}^{(T)}(\omega_k^{(T)})} \right) + \left| E[d_{\underline{X}}^{(T)}(\omega_k^{(T)})] \right|^2. \quad (2.7)$$

From Proposition 2.1

$$T^{-1}\mathrm{Cov}\left( d_{\underline{X}}^{(T)}(\omega_k^{(T)}), \overline{d_{\underline{X}}^{(T)}(\omega_k^{(T)})} \right) = T^{-1}\mathrm{Cov}\left( d_{\underline{X}}^{(T)}(\omega_k^{(T)}), d_{\underline{X}}^{(T)}(-\omega_k^{(T)}) \right)$$

$$= T^{-1}\mathrm{Cov}\left( d_{\underline{X}^{(B)}}^{(\tau)}(\omega_k^{(T)}) + e^{i\tau\omega_k^{(\tau)}} d_{\underline{X}^{(A)}}^{(T-\tau)}(\omega_k^{(T)}), d_{\underline{X}^{(B)}}^{(\tau)}(-\omega_k^{(T)}) + e^{-i\tau\omega_k^{(\tau)}} d_{\underline{X}^{(A)}}^{(T-\tau)}(-\omega_k^{(T)}) \right)$$

$$\sim 2\pi\lambda f_B(\omega_k^{(T)}) + 2\pi(1-\lambda) f_A(\omega_k^{(T)}),$$

where the orthonormal basis is partitioned as $\underline{e}_T(\omega_k^{(T)}) = \left( \underline{e}'_\tau(\omega_k^{(T)}), e^{i\tau\omega_k^{(T)}} \underline{e}'_{T-\tau}(\omega_k^{(T)}) \right)'$. Note that, for $k \neq 0 \mod(T)$, we have that

$$0 = \langle \underline{e}_T(0), \underline{e}_T(\omega_k^{(T)}) \rangle = \langle \underline{1}, \underline{e}_T(\omega_k^{(T)}) \rangle = \langle \underline{1}, \underline{e}_\tau(\omega_k^{(T)}) \rangle + e^{i\tau\omega_k^{(T)}} \langle \underline{1}, \underline{e}_{T-\tau}(\omega_k^{(T)}) \rangle,$$

where $\underline{1}$ denotes a vector of 1's with its appropriate dimension. Since $e^{i\tau\omega_k^{(T)}} \langle \underline{1}, \underline{e}_{T-\tau}(\omega_k^{(T)}) \rangle = -\langle \underline{1}, \underline{e}_\tau(\omega_k^{(T)}) \rangle$, it follows that

$$T^{-1}\left| E[d_{\underline{X}}^{(T)}(\omega_k^{(T)})] \right|^2 = (\mu_B - \mu_A)^2 \frac{1}{T} \left| \frac{1 - e^{i\tau\omega_k^{(T)}/2}}{1 - e^{i\omega_k^{(T)}/2}} \right|^2 = (\mu_B - \mu_A)^2 \frac{\sin^2(\tau \omega_k^{(T)}/2)}{T \sin^2(\omega_k^{(T)}/2)}.$$

This completes the proof of the theorem. □

***Remarks***.

1. It is noted that for zero frequency, $k = 0$,

$$E[d_{\underline{X}}^{(T)}(0)] = \langle \underline{1}, E[\underline{X}_T] \rangle = \langle \underline{1}, \mu_B \underline{1}_{1:\tau} + \mu_A \underline{1}_{\tau+1:T} \rangle = T(\lambda_T \mu_B + (1-\lambda_T)\mu_A),$$

which follows that $\left| E[d_{\underline{X}}^{(T)}(0)] \right|^2 = T^2 \{\lambda_T \mu_B + (1-\lambda_T)\mu_A\}^2$.

2. If $X_0, X_1, \cdots, X_{T-1}$ is a restricted independently and identically distributed (i.i.d.) time ordered sequence, it is known that $f_s(\omega) = \sigma_s^2/2\pi$, $s = B, A$, then

$$T^{-1} E[I_{\underline{X}}^{(T)}(0)] = T^{-1}(\lambda_T \sigma_B^2 + (1-\lambda_T)\sigma_A^2) + (\lambda_T \mu_B + (1-\lambda_T)\mu_A)^2,$$

which shows that

$$\lim_{T \to \infty} T^{-1} E[I_{\underline{X}}^{(T)}(0)] = (\lambda \mu_B + (1-\lambda)\mu_A)^2.$$

3. It should be pointed out that the time horizon $T$ is sufficiently large. In addition, $\lambda_T$ remains proportional to $T$. This condition implies that there exists sufficient information before and after the change-point for estimating the parameters involved. This is exactly the same requirement under the time domain (see e.g., Fotopoulos et al. 2010).

4. If there is no change-point, then obviously,

$$T^{-1} Cov(d_{\underline{X}}^{(T)}(\omega_k^{(T)}), d_{\underline{X}}^{(T)}(\omega_k^{(T)})) \sim 2\pi f(\omega_k^{(T)}).$$

5. For $k \neq 0 \bmod (T)$ in $\omega_k^{(T)} = 2\pi k/T$, and assuming that $X_0, X_1, \cdots, X_{T-1}$ is an i.i.d. time ordered sequence, the expected periodogram under the change-point scenario satisfies

$$E[I_{\underline{X}}^{(T)}(\omega_k^{(T)})] \sim \lambda \sigma_B^2 + (1-\lambda)\sigma_A^2 + (\mu_B - \mu_A)^2 \frac{\sin^2(\lambda_T \pi k)}{T \sin^2(\omega_k/2)}.$$

6. In contrast to Yau and Davis (2012) whom maintain only the frequencies around the zero necessary for their analysis (discrimination between CP versus LRD), this study requires all possible frequencies.

To compute the covariance of the periodogram, cumulants of order four of the discrete Fourier transform are essential (see e.g., Terdik, 1999). We begin implementing a few general results. Let $X = (X^{(1)}, \cdots, X^{(k)})$ be a $k$-dimensional random vector. The first three moments of the components are defined as follows:

$k^{(1)} = E[X^{(1)}]$ for the components of the mean vector,

$k^{(1)(2)} = E[X^{(1)} X^{(2)}]$ for the components of the second moment matrix,

$k^{(1)(2)(3)} = E[X^{(1)} X^{(2)} X^{(3)}]$ for the third moments, and so on.

Next, Einstein's summation convention is applied, i.e., $\xi_{(1)} X^{(1)}$ denotes the linear combination $\xi_{(1)} X^{(1)} + \cdots + \xi_{(k)} X^{(k)}$, the square of the linear combination is $(\xi_{(1)} X^{(1)})^2 = \xi_{(1)} \xi_{(2)} X^{(1)} X^{(2)}$, i.e., a sum of $k^2$, and so on for higher powers. The Taylor expansion of the moment generating function $M(\xi) = E[\exp(\xi_{(1)} X^{(1)})]$ provides

$$M(\xi) = 1 + \xi_{(1)} k^{(1)} + \frac{1}{2!}\xi_{(1)}\xi_{(2)} k^{(1)(2)} + \frac{1}{3!}\xi_{(1)}\xi_{(2)}\xi_{(3)} k^{(1)(2)(3)} + \cdots.$$

Similarly, the cumulants $k^{(1),(2)}, k^{(1),(2),(3)}, \cdots$ can be obtained from $\ln(M(\xi))$, in expanding $\ln(M(\xi))$ in Taylor series as

$$\ln(M(\xi)) = \xi_{(1)} k^{(1)} + \frac{1}{2!}\xi_{(1)}\xi_{(2)} k^{(1),(2)} + \frac{1}{3!}\xi_{(1)}\xi_{(2)}\xi_{(3)} k^{(1),(2),(3)} + \cdots.$$

Equating the coefficients reveal that each moment $k^{(1)(2)}, k^{(1)(2)(3)}, \cdots$ is a sum over partitions of the superscript, each term in the sum being a product of cumulants

$$k^{(1)(2)} = k^{(1),(2)} + k^{(1)} k^{(2)},$$

$$k^{(1)(2)(3)} = k^{(1),(2),(3)} + k^{(1),(2)} k^{(3)} + k^{(1),(2)} k^{(3)} + k^{(3),(2)} k^{(1)} + k^{(1)} k^{(2)} k^{(3)} k^{(3)}$$

$$= k^{(1),(2),(3)} + k^{(1),(2)} k^{(3)}[3] + k^{(1)} k^{(2)},$$

$$k^{(1)(2)(3)(4)} = k^{(1),(2),(3),(4)} + k^{(1),(2),(3)} k^{(4)}[4] + k^{(1),(2)} k^{(3),(4)}[3] + k^{(1),(2)} k^{(3)} k^{(4)}[6] + k^{(1)} k^{(2)} k^{(3)} k^{(4)}.$$

Each parenthetical number indicates a sum over distinct partitions having the same block sizes. What is required, however, in the analysis of the periodogram is the component $k^{(1)(2),(3)(4)}$. As above, we express this term in terms of the kurtosis, skewness, covariances and the first moments as

$$k^{(1)(2),(3)(4)} = k^{(1),(2),(3),(4)} + k^{(1),(2),(3)} k^{(4)}[4] + k^{(1),(3)} k^{(2),(4)} + k^{(1),(4)} k^{(2),(3)}$$

$$+ k^{(1)} k^{(3)} k^{(2),(4)} + k^{(1)} k^{(4)} k^{(2),(3)} + k^{(2)} k^{(3)} k^{(1),(4)} + k^{(2)} k^{(4)} k^{(1),(3)}. \qquad (2.8)$$

In light of (2.8), the variance-covariance of the periodogram is represented in the next Theorem.

**Theorem 2.2.** *Let $\lambda_T \in (0,1)$ and let the two successive segments $\underline{X}_B = (X_0, X_1, \cdots, X_{\tau-1})$ and $\underline{X}_A = (X_\tau, X_{\tau+1}, \cdots, X_{T-1})$ be in $\mathscr{G}$. Further, let the two independent segments consist of i.i.d. random variables with $E[X_i] = \mu_B$, $Var(X_i) = \sigma_B^2$, for $i \leq \tau - 1$, and $E[X_i] = \mu_A$, $Var(X_i) = \sigma_A^2$, for $i \geq \tau$, respectively. Then, the variance-covariance of the periodogram satisfies the following:*

1. *For $\omega_k, \omega_j \in [0, 2\pi]$ that satisfy $\omega_k \pm \omega_j \neq 0 \mod(2\pi)$ and for subsequences $j_T, k_T \neq 0 \mod(T)$ that $\lim_{T \to \infty} \omega_{k_T}^{(T)} = \omega_k$ and $\lim_{T \to \infty} \omega_{j_T}^{(T)} = \omega_j$, the periodogram satisfies*

$$\lim_{T \to \infty} Cov\left(I_{\underline{X}}^{(T)}(\omega_{k_T}^{(T)}), I_{\underline{X}}^{(T)}(\omega_{j_T}^{(T)})\right) = 0.$$

2. *For $\omega_k \pm \omega_j = 0 \mod(2\pi)$ and $\omega_k \neq 0 \mod(2\pi)$, then*

$$\lim_{T\to\infty} Cov\left(I_{\underline{X}}^{(T)}(\omega_k^{(T)}), I_{\underline{X}}^{(T)}(\omega_j^{(T)})\right) = \left(\lambda \sigma_B^2 + (1-\lambda)\sigma_A^2\right)^2.$$

**Proof.** Note that the periodogram is expressed in terms of the Fourier transforms as

$$T^2 Cov\left(I_{\underline{X}}^{(T)}(\omega_k), I_{\underline{X}}^{(T)}(\omega_j)\right) = Cum\left(d_{\underline{X}}^{(T)}(\omega_k) d_{\underline{X}}^{(T)}(-\omega_k), d_{\underline{X}}^{(T)}(\omega_j) d_{\underline{X}}^{(T)}(-\omega_j)\right).$$

Thus, for $\omega_k^{(T)} \neq 0 \bmod (2\pi)$, the orthogonal property of $e_T(\omega_k^{(T)})$, for $\omega_k^{(T)} \in [0, 2\pi]$, and for $\lambda_T \in (0,1)$ the sequence $X_0, X_1, \cdots, X_{T-1}$ is fragmented into two independent i.i.d. sequences, in which case the first moment of the periodogram is formed as

$$k^{(1)} = E\left(d_X^{(T)}(\omega_{k_T}^{(T)})\right) = \sum_{t=0}^{T-1} e^{it\omega_{k_T}^{(T)}} E[X_t] = \mu_B \left\langle 1, e_\tau(\omega_{k_T}^{(T)})\right\rangle + \mu_A e^{i\tau_T(\omega_{k_T}^{(T)})}\left\langle 1, e_{T-\tau}(\omega_{k_T}^{(T)})\right\rangle$$

$$\sim (\mu_B - \mu_A)\langle 1, \underline{e}_\tau(\omega_k)\rangle. \tag{2.9}$$

Since Lemma 2.1, the right hand side of (2.9), is bounded by a constant.

Also, for $\omega_k^{(T)} \pm \omega_j^{(T)} \neq 0 \bmod(2\pi)$, the orthogonal property of $e_T(\omega_k^{(T)})$, and for $\omega_k^{(T)} \in [0, 2\pi]$, similar arguments as in (2,9), provide the following

$$k^{(1),(3)} = Cov\left(d_X^{(T)}(\omega_{k_T}^{(T)}), d_X^{(T)}(\omega_{j_T}^{(T)})\right) = Cov\left(\left\langle e_T(\omega_{k_T}^{(T)}), \underline{X}_{1:T}\right\rangle, \left\langle e_T(\omega_{j_T}^{(T)}), \underline{X}_{1:T}\right\rangle\right)$$

$$= \sum_{s,t=0}^{T} e^{i\left(t\omega_{k_T}^{(T)} + s\omega_{j_T}^{(T)}\right)} Cov(X_s, X_t)$$

$$= \sigma_B^2 \left\langle 1, \underline{e}_\tau(\omega_{k_T}^{(T)} + \omega_{j_T}^{(T)})\right\rangle + \sigma_A^2 e^{i\tau_T(\omega_{k_T}^{(T)} + \omega_{j_T}^{(T)})}\left\langle 1, \underline{e}_{T-\tau}(\omega_{k_T}^{(T)} + \omega_{j_T}^{(T)})\right\rangle$$

$$\sim (\sigma_B^2 - \sigma_A^2)\langle 1, \underline{e}_\tau(\omega_k + \omega_j)\rangle. \tag{2.10}$$

Next, from Lemma 2.1 $|\langle 1, \underline{e}_\tau(\omega_k + \omega_j)\rangle| = 1/|\sin((\omega_k + \omega_j)/2)|$, $k^{(1),(3)}$ is also bounded by a constant. The same applies for combinations $(1),(4)$, $(2),(3)$, and $(2),(4)$.

Similarly, for $\omega_k, \omega_k \neq 0 \bmod (2\pi)$ and assuming that for $\lambda_T \in (0,1)$ the skewness before and after the change point $\tau$ satisfies $\varsigma_B = Cum(X_1, X_1, X_1) \neq Cum(X_T, X_T, X_T) = \varsigma_A$. Then, under the i.i.d. case, the skewness of the Fourier transform satisfies

$$k^{(1),(2),(3)} = cum\left(d_X^{(T)}(\omega_{k_T}^{(T)}), d_X^{(T)}(-\omega_{k_T}^{(T)}), d_X^{(T)}(\omega_{j_T}^{(T)})\right)$$

$$= \sum_{t,s,h=1}^{T} e^{i\left(t\omega_{k_T}^{(T)} - s\omega_{k_T}^{(T)} + h\omega_{j_T}^{(T)}\right)} Cum(X_t, X_s, X_h)$$

$$= \varsigma_B \left\langle 1, \underline{e}_\tau(\omega_{j_T}^{(T)})\right\rangle + \varsigma_A e^{i\tau\omega_{j_T}^{(T)}}\left\langle 1, \underline{e}_{T-\tau}(\omega_{j_T}^{(T)})\right\rangle$$

$$\sim (\varsigma_B - \varsigma_A)\langle 1, \underline{e}_\tau(\omega_j)\rangle. \tag{2.11}$$

Again, (2.11) is bounded the same way as in (2.9). Obviously, when $\zeta_B = \zeta_A$, $k^{(1),(2),(3)} = 0$. The same arguments apply for any other triplet combinations of the superscripts $(j)$, $j = 1, 2, 3, 4$.

Finally, under the two segments and since each one of them consist of i.i.d. observations, the kurtoses satisfy

$$k_B = Cum(X_1, X_1, X_1, X_1) \neq Cum(X_T, X_T, X_T, X_T) = k_A,$$

the fourth cumulants of the Fourier transform is expressed as

$$k^{(1),(2),(3),(4)} = cum\left(d_X^{(T)}(\omega_{k_T}^{(T)}), d_X^{(T)}(-\omega_{k_T}^{(T)}), d_X^{(T)}(\omega_{j_T}^{(T)}), d_X^{(T)}(-\omega_{j_T}^{(T)})\right)$$
$$\sim T\{\lambda k_B + (1-\lambda)k_A\}. \tag{2.12}$$

Collecting (2.9)-(2.12), it can be seen that for $\omega_k^{(T)}, \omega_j^{(T)} \neq 0 \mod(2\pi)$ and $\omega_k^{(T)} \pm \omega_j^{(T)} \neq 0 \mod(2\pi)$

$$\lim_{T\to\infty} Cov\left(I_{\underline{X}}^{(T)}(\omega_{k_T}^{(T)}), I_{\underline{X}}^{(T)}(\omega_{j_T}^{(T)})\right) = \lim_{T\to\infty} T^{-2} Cum\left(d_X^{(T)}(\omega_{k_T}^{(T)})d_X^{(T)}(-\omega_{k_T}^{(T)}), d_X^{(T)}(\omega_{j_T}^{(T)})d_X^{(T)}(-\omega_{j_T}^{(T)})\right) = 0.$$

This is because all terms on the right hand side of (2.8) are either bounded by constants or they are of smaller order than $T^2$.

For the case $\omega_k = \omega_j$ and $\omega_k \neq 0 \mod(2\pi)$, one applies the same method as earlier. The only term that is of order $T^2$ is $k^{(1),(4)}k^{(2),(3)}$, the remaining terms are either bounded by a constant or they are of smaller order than $T^2$. To see this we only consider the term that is of order $T^2$, i.e., we have

$$k^{(1),(4)}k^{(2),(3)} \sim T^2 \left\{\lambda \sigma_B^2 + (1-\lambda)\sigma_A^2\right\}^2.$$

This completes the proof of Theorem 2.2. □

**Remark.** If $\lambda_T = 1$, i.e., there is no change-point, then the convergence to the true limit is much faster. Under this scenario, several terms are zero.

## 3. Parameter estimation and Gauss-Newton methodology

To develop an estimation methodology for the change-point problem using frequency domain, we assume that there exist two successive segments $\underline{X}_B = (X_0, X_1, \cdots, X_{\tau-1})$ and $\underline{X}_A = (X_\tau, X_{\tau+1}, \cdots, X_{T-1})$ in $\mathscr{G}$ such that $E[X_i] = \mu_B$, $Var(X_i) = \sigma_B^2$, for $i \leq \tau - 1$, and $E[X_i] = \mu_A$, $Var(X_i) = \sigma_A^2$, for $i \geq \tau$. We next define the parameters $\sigma^2, \mu^2$ as $\sigma^2 = \lambda \sigma_B^2 + (1-\lambda)\sigma_A^2$ and $\mu^2 = (\mu_B - \mu_A)^2$, where $\lambda$ is the normalized change-point. Based on the frequency domain methodology, one observes the periodogram at the discrete frequencies $\omega_k^{(T)} = 2\pi k/T$, $k = 0, 1, \cdots, \lfloor T/2 \rfloor$. Write $\underline{\omega}_T = (2\pi/T, \cdots, 2\pi\lfloor T/2 \rfloor/T)$. The periodogram vector is then expressed as $I_{\underline{X}}^{(T)}(\underline{\omega}_T) = \left(I_{\underline{X}}^{(T)}(\omega_1^{(T)}), \cdots, I_{\underline{X}}^{(T)}(\omega_{T/2}^{(T)})\right)'$. In addition, introduce the vector of ones $\underline{1} = (1,\cdots,1)'$, the nonlinear vector $\underline{g}_T(\lambda, \underline{\omega}_T) = \left(g_{1,T}(\lambda, \omega_1^{(T)}), \cdots, g_{T/2,T}(\lambda, \omega_{T/2}^{(T)})\right)'$ and an unobservable error vector $\underline{u}_T = (u_{T,1}, \cdots, u_{T,T/2})$. The vector $\underline{g}_T(\lambda, \underline{\omega}_T)$ is defined as follows

$$\underline{g}_T(\lambda, \underline{\omega}_T) \equiv \left(\frac{\sin^2(\lambda\pi)}{T\sin^2(\pi/T)}, \cdots, \frac{\sin^2(\lambda(T/2)\pi)}{T\sin^2(\pi(T/2)/T)}\right).$$

From Theorem 2.1, we have that $\underline{u}_T = I_{\underline{X}}^{(T)}(\underline{\omega}_T) - \sigma^2 \underline{1} - \mu^2 \underline{g}(\lambda, \underline{\omega}_T)$, and in conjunction with Theorem 2.2, $\lim_{T\to\infty} Var(\underline{u}_T) = \sigma^4 I$, where $I$ is the identity matrix. Note that the dimension of the vectors defined above is of order $\lfloor T/2 \rfloor$. Due to the symmetry around $\lfloor T/2 \rfloor$, we only consider half of the periodogram values, i.e., $\frac{\sin^2(\lambda_T k\pi)}{T\sin^2(\pi k/T)} = \frac{\sin^2(\lambda_T (T-k)\pi)}{T\sin^2(\pi(T-k)/T)}$, $k = 1,\cdots,\lfloor T/2 \rfloor$. Conveniently, rewrite the periodogram in a vector form as:

$$I_{\underline{X}}^{(T)}(\underline{\omega}_T) = \underline{f}(\underline{\omega}_T, \underline{\theta}) + \underline{u}_T, \qquad (3.1)$$

where $\underline{f}(\underline{\omega}_T, \underline{\theta}) = \sigma^2 \underline{1} + \mu^2 \underline{g}(\lambda, \underline{\omega}_T)$ and $\underline{\theta} = (\sigma^2, \mu^2, \lambda)'$. Note that $f(2\pi k/T, \underline{\theta}) = \sigma^2 + \mu^2 g(\lambda, 2\pi k/T)$. Set $f(k,\underline{\theta}) = f(2\pi k/T, \underline{\theta})$. $f(k,\underline{\theta}): \mathbf{R} \times \Theta \to \mathbf{R}$ is deterministic on $\mathbf{R}$ for each $\underline{\theta} \in \Theta$ and continuous on $\Theta$, with $(\sigma^2, \mu^2) \in \mathbf{R}_+^2$ and $\lambda \in (0, 1/2)$. The parameter $\underline{\theta}$ is an unknown $3 \times 1$ vector to be estimated. Although, the unobservable array of errors $u_{T,1},\cdots,u_{T,T/2}$ is asymptotic independent, some of the classical results of the estimation theory still hold (see e.g., White and Domowitz, 1984). In particular, the existence of $\hat{\underline{\theta}}_T$ is ensured by Lemma 2 of Jennrich (1969), since the functional form of $f_k = f(2\pi k/T, \cdot)$ satisfies all the conditions of Lemma 2. The nonlinear least squares estimator $\hat{\underline{\theta}}_T$ is then obtained using the quadratic function

$$S_T(\underline{\theta}) = \left(I_{\underline{X}}^{(T)}(\underline{\omega}_T) - \sigma^2 \underline{1} - \mu^2 \underline{g}_T(\lambda, \underline{\omega}_T)\right)'\left(I_{\underline{X}}^{(T)}(\underline{\omega}_T) - \sigma^2 \underline{1} - \mu^2 \underline{g}_T(\lambda, \underline{\omega}_T)\right), \qquad (3.2)$$

in which case the estimator $\hat{\underline{\theta}}_T$ is $\hat{\underline{\theta}}_T = \arg\min_{\theta \in \Theta} S_T(\theta)$. More accurately, one could consider a weighted least squared estimator of the $\hat{\underline{\theta}}_T$, which can be attained by simply minimizing

$$S_T^*(\underline{\theta}) = \left(I_{\underline{X}}^{(T)}(\underline{\omega}_T) - \sigma^2 \underline{1} - \mu^2 \underline{g}_T(\lambda, \underline{\omega}_T)\right)' \hat{V}_T^{-1} \left(I_{\underline{X}}^{(T)}(\underline{\omega}_T) - \sigma^2 \underline{1} - \mu^2 \underline{g}_T(\lambda, \underline{\omega}_T)\right),$$

where $\hat{V}_T$ is a consistent estimator of the asymptotic variance-covariance matrix of $\underline{u}_T$. On the other hand, since $\lim_{T\to\infty} Var(\underline{u}_T) = \sigma^4 I$, being constant, and $\sigma^2$ is the same parameter as in (3.1), it is preferable to consider $S_T(\theta)$ as in (3.2).

To continue, one may note that an estimate of $\lambda$ satisfies $\lambda \neq \pm 1 \bmod (1)$. Under this, one cannot make a distinction between $\lambda \in (0,1/2)$ and $\lambda \in (1/2,1)$. The identification of whether $\lambda$ lies in $(0,1/2)$ or $(1/2,1)$ will be computed after an estimator of $\sigma^2$ is available. The procedure suggested will always provide estimates of $\lambda \in (0,1/2)$. As soon as $\hat{\lambda}$ is evaluated, one then estimates both the pairs $(\hat{\sigma}_B^2(\hat{\lambda}), \hat{\sigma}_A^2(\hat{\lambda}))$ and $(\hat{\sigma}_B^2(1-\hat{\lambda}), \hat{\sigma}_A^2(1-\hat{\lambda}))$. For each pair, we compute $\hat{\sigma}^2(\hat{\lambda})$ and $\hat{\sigma}^2(1-\hat{\lambda})$. Then, whenever $\hat{\sigma}^2(\hat{\lambda}) < \hat{\sigma}^2(1-\hat{\lambda})$, $\hat{\lambda} \in (0,1/2)$ and when $\hat{\sigma}^2(\hat{\lambda}) > \hat{\sigma}^2(1-\hat{\lambda})$, $\hat{\lambda} \in (1/2,1)$.

To establish an estimate of $\underline{\theta}$, an iteration method is formulated. Towards this, we simply update the estimated values of $\hat{\underline{\theta}}_T$ by adding the scaled gradient at each step. Specifically, $\hat{\underline{\theta}}_T$ is attained using the Gauss-Newton algorithm as follows:

$$\hat{\underline{\theta}}_k = \hat{\underline{\theta}}_{k-1} - \left(\nabla^2 S_T(\hat{\underline{\theta}}_{k-1})\right)^{-1} \nabla S_T(\hat{\underline{\theta}}_{k-1}), \tag{3.3}$$

where the gradient and Hessian are defined as follows

$$\nabla S_T(\underline{\theta}) = \frac{\partial S_T(\underline{\theta})}{\partial \underline{\theta}} = -2\langle J_T, \underline{u}_T \rangle, \text{ and} \tag{3.4}$$

$$H = \frac{\partial^2 S_T(\underline{\theta})}{\partial \underline{\theta} \, \partial \underline{\theta}} = \nabla^2 S_T(\underline{\theta}) = 2\langle J_T, J_T \rangle - 2\sum_{j=0}^{T-1} u_{T,j} \left( \frac{\partial^2 g(\underline{\theta}, \omega_j^{(T)})}{\partial \underline{\theta} \, \partial \underline{\theta}} \right). \tag{3.5}$$

Without violating the notation, $J_T = \left( \frac{\partial f(k,\underline{\theta})}{\partial \underline{\theta}} \right)_{T/2 \times 3}$ represents the Jacobian matrix at iteration $k$.

From (3.2), the parameters $\sigma^2$ and $\mu^2$ are in a linear form. In this case, the parameters $\sigma^2$ and $\mu^2$ can be computed in a much simpler manner. However, for the parameter $\lambda$ a Gauss-Newton approach is applied. In particular, equations (3.3)-(3.5) are now modified as

$$\hat{\sigma}_k^2 = \left(\frac{T}{2}-1\right)^{-1} \left\langle \underline{1}, \underline{I}_{\underline{X}}^{(T)}(\underline{\omega}_T) - \hat{\mu}_{k-1}^2 \underline{g}(\hat{\lambda}_{k-1}, \underline{\omega}_T) \right\rangle, \tag{3.6}$$

$$\hat{\mu}_k^2 = \frac{\left\langle \underline{g}(\hat{\lambda}_{k-1}, \underline{\omega}_T), \underline{I}_{\underline{X}}^{(T)}(\underline{\omega}_T) - \hat{\sigma}_{k-1}^2 \underline{1} - \hat{\mu}_{k-1}^2 \underline{g}(\hat{\lambda}_{k-1}, \underline{\omega}_T) \right\rangle}{\left\langle \underline{g}(\hat{\lambda}_{k-1}, \underline{\omega}_T), \underline{g}(\hat{\lambda}_{k-1}, \underline{\omega}_T) \right\rangle}, \tag{3.7}$$

$$\hat{\lambda}_k = \hat{\lambda}_{k-1} + \frac{\left\langle \left. \frac{\partial \underline{g}(\lambda,\underline{\omega}_T)}{\partial \lambda} \right|_{\lambda=\hat{\lambda}_{k-1}}, \underline{I}_X^{(T)}(\underline{\omega}_T) - \hat{\sigma}_{k-1}^2 \underline{1} - \hat{\mu}_{k-1}^2 \underline{g}(\hat{\lambda}_{k-1},\underline{\omega}_T) \right\rangle}{\hat{\mu}_{k-1}^2 \left\langle \left. \frac{\partial \underline{g}(\lambda,\underline{\omega}_T)}{\partial \lambda} \right|_{\lambda=\hat{\lambda}_{k-1}}, \left. \frac{\partial \underline{g}(\lambda,\underline{\omega}_T)}{\partial \lambda} \right|_{\lambda=\hat{\lambda}_{k-1}} \right\rangle}. \qquad (3.8)$$

Again, when $\hat{\lambda}_{k-1}$ is given, we obtain $\hat{\sigma}_k^2$ and $\hat{\mu}_k^2$ in a linear manner. Conversely, one uses the Gauss-Newton method to evaluate $\hat{\lambda}_k$ with given $\hat{\sigma}_{k-1}^2$ and $\hat{\mu}_{k-1}^2$.

For the asymptotic variance-covariance of the estimator $\hat{\theta}_T$, the inverse of the matrix $J_T' J_T$ is required. At first, let the Jacobian matrix be as follows;

$$J_T = \left( \frac{\partial f(k,\underline{\theta})}{\partial \underline{\theta}} \right)_{T/2 \times 3} = \left( 1 \quad \underline{g}(\lambda,\underline{\omega}_T) \quad \mu^2 \frac{\partial \underline{g}(\lambda,\underline{\omega}_T)}{\partial \lambda} \right). \qquad (3.9)$$

Thus,

$$J_T' J_T = \begin{pmatrix} \langle 1, 1 \rangle & \langle 1, \underline{g}(\lambda,\underline{\omega}_T) \rangle & \mu^2 \left\langle 1, \frac{\partial \underline{g}(\lambda,\underline{\omega}_T)}{\partial \lambda} \right\rangle \\ \langle 1, \underline{g}(\lambda,\underline{\omega}_T) \rangle & \langle \underline{g}(\lambda,\underline{\omega}_T), \underline{g}(\lambda,\underline{\omega}_T) \rangle & \mu^2 \left\langle \underline{g}(\lambda,\underline{\omega}_T), \frac{\partial \underline{g}(\lambda,\underline{\omega}_T)}{\partial \lambda} \right\rangle \\ \mu^2 \left\langle 1, \frac{\partial \underline{g}(\lambda,\underline{\omega}_T)}{\partial \lambda} \right\rangle & \mu^2 \left\langle \underline{g}(\lambda,\underline{\omega}_T), \frac{\partial \underline{g}(\lambda,\underline{\omega}_T)}{\partial \lambda} \right\rangle & \mu^4 \left\langle \frac{\partial \underline{g}(\lambda,\underline{\omega}_T)}{\partial \lambda}, \frac{\partial \underline{g}(\lambda,\underline{\omega}_T)}{\partial \lambda} \right\rangle \end{pmatrix}. \qquad (3.10)$$

Upon using Lemma A.1 and Lemma A.2 (see Appendix) the asymptotic $J_T' J_T$ for $\lambda \in (0, 1/2)$ is expressed as

$$J_T' J_T = \begin{pmatrix} T/2 & \frac{\lambda(1-\lambda)}{2} T + O(1) & \mu^2 \left( \frac{1}{2} - \lambda \right) T + O(1) \\ \frac{\lambda(1-\lambda)}{2} T + O(1) & \frac{1}{2} \left( \frac{2\lambda^3}{3} - \lambda^4 \right) T^2 + O(T) & \frac{\mu^2}{2} \left( \lambda^2 - 2\lambda^3 \right) T^2 + O(T) \\ \mu^2 \left( \frac{1}{2} - \lambda \right) T + O(1) & \frac{\mu^2}{2} \left( \lambda^2 - 2\lambda^3 \right) T^2 + O(T) & \mu^4 \left( \lambda - 2\lambda^2 \right) T^2 + O(T) \end{pmatrix} \qquad (3.11)$$

Further, from (3.11), and Fuller (pp. 260-266, 2009), the main contributions of this article are summarized in the following two theorems.

**Theorem 3.1.** *Assume that model (3.1) holds. Let* $E[X^4] < \infty$. *If*

$$(\sigma_0^2 - \sigma^2) + \mu_0^2 (\lambda_0 - \lambda_0^2) - \mu^2 (\lambda - \lambda^2) \neq 0,$$

*then the least square estimate* $\hat{\underline{\theta}}_T$ *is a consistent estimator of* $\underline{\theta}_0 \in \Theta \subseteq \mathbf{R}_+^2 \times (0,1/2)$.

**Proof.** As in Jennrich (1969), write $k_T(\underline{\theta},\underline{\theta}_0) = \langle \underline{f}(\omega_T,\underline{\theta}_0) - \underline{f}(\omega_T,\underline{\theta}), \underline{f}(\omega_T,\underline{\theta}_0) - \underline{f}(\omega_T,\underline{\theta}) \rangle$, where $\underline{\theta}_0 = (\sigma_0^2, \mu_0^2, \lambda_0)'$ belongs to $\Theta \subseteq \mathbf{R}_+^2 \times (0,1/2)$. Again, $f(k,\underline{\theta}) = \sigma^2 + \mu^2 g(\lambda,k)$ are continuous functions in $\underline{\theta} \in \Theta$. To show consistency, it is required to show that $\lim_{T \to \infty} k_T(\underline{\theta},\underline{\theta}_0) = \infty$, for all $\underline{\theta} \neq \underline{\theta}_0$. The strategy applied here is to show weak consistency as a modification of Lemma 1 in Wu (1981). Specifically, it can be seen that

$$S_T(\underline{\theta}) - S_T(\underline{\theta}_0) = \langle I_{\underline{X}}^{(T)}(\omega_T) - \underline{f}(\omega_T,\underline{\theta}), I_{\underline{X}}^{(T)}(\omega_T) - \underline{f}(\omega_T,\underline{\theta}) \rangle - \langle I_{\underline{X}}^{(T)}(\omega_T) - \underline{f}(\omega_T,\underline{\theta}_0), I_{\underline{X}}^{(T)}(\omega_T) - \underline{f}(\omega_T,\underline{\theta}_0) \rangle$$

$$= \langle \underline{f}(\omega_T,\underline{\theta}_0) - \underline{f}(\omega_T,\underline{\theta}), \underline{f}(\omega_T,\underline{\theta}_0) - \underline{f}(\omega_T,\underline{\theta}) \rangle - 2\langle \underline{f}(\omega_T,\underline{\theta}_0) - \underline{f}(\omega_T,\underline{\theta}), \underline{u}_T \rangle$$

$$= k_T(\underline{\theta},\underline{\theta}_0) \left\{ 1 - 2 \frac{\langle \underline{f}(\omega_T,\underline{\theta}_0) - \underline{f}(\omega_T,\underline{\theta}), \underline{u}_T \rangle}{k_T(\underline{\theta},\underline{\theta}_0)} \right\}. \tag{3.12}$$

In order for $\hat{\underline{\theta}}_T$ to be weakly consistent, it suffices to show that $\underline{\lim}_{T \to \infty} k_T(\underline{\theta},\underline{\theta}_0) = \infty$, for all $\underline{\theta} \neq \underline{\theta}_0$, and then to prove that

$$\frac{\langle \underline{f}(\omega_T,\underline{\theta}_0) - \underline{f}(\omega_T,\underline{\theta}), \underline{u}_T \rangle}{k_T(\underline{\theta},\underline{\theta}_0)} \to 0 \text{ in probability.} \tag{3.13}$$

From the Cauchy-Swartz inequality, it can be seen that

$$k_T(\underline{\theta},\underline{\theta}_0) = \sum_{k=0}^{\lfloor T/2 \rfloor} \left\{ (\sigma_0^2 - \sigma^2) + (\mu_0^2 - \mu^2) \frac{\sin^2(\pi k \lambda)}{T \sin^2(\pi k/T)} + \mu_0^2 \frac{\sin^2(\pi k \lambda_0) - \sin^2(\pi k \lambda)}{T \sin^2(\pi k/T)} \right\}^2$$

$$\geq \frac{1}{\lfloor T/2 \rfloor} \left\{ \sum_{k=0}^{\lfloor T/2 \rfloor} \left\{ (\sigma_0^2 - \sigma^2) + (\mu_0^2 - \mu^2) \frac{\sin^2(\pi k \lambda)}{T \sin^2(\pi k/T)} + \mu_0^2 \frac{\sin^2(\pi k \lambda_0) - \sin^2(\pi k \lambda)}{T \sin^2(\pi k/T)} \right\} \right\}^2$$

$$= \frac{1}{\lfloor T/2 \rfloor} \left\{ \sum_{k=0}^{\lfloor T/2 \rfloor} (\sigma_0^2 - \sigma^2) + T \left\{ (\mu_0^2 - \mu^2) \left( \frac{\lambda - \lambda^2}{2} \right) + \mu_0^2 \left( \frac{\lambda_0 - \lambda_0^2}{2} - \frac{\lambda - \lambda^2}{2} \right) + O(T^{-1}) \right\} \right\}^2$$

$$= T \left\{ (\sigma_0^2 - \sigma^2) + \mu_0^2 (\lambda_0 - \lambda_0^2) - \mu^2 (\lambda - \lambda^2) + O(T^{-1}) \right\}^2. \tag{3.14}$$

Thus, for $(\sigma_0^2 - \sigma^2) + \mu_0^2(\lambda_0 - \lambda_0^2) - \mu^2(\lambda - \lambda^2) \neq 0$, $T^{-1} k_T(\underline{\theta},\underline{\theta}_0) > 0$. This, in turn, also shows that $\underline{\lim}_{T \to \infty} k_T(\underline{\theta},\underline{\theta}_0) = \infty$.

Next, applying Chebyshev's inequality, we have that, for $\delta > 0$,

$$P\left( \frac{|\langle \underline{f}(\omega_T,\underline{\theta}_0) - \underline{f}(\omega_T,\underline{\theta}), \underline{u}_T \rangle|}{k_T(\underline{\theta},\underline{\theta}_0)} > \delta \right) \leq \frac{\text{Var}(\langle \underline{f}(\omega_T,\underline{\theta}_0) - \underline{f}(\omega_T,\underline{\theta}), \underline{u}_T \rangle)}{\delta^2 k_T^2(\underline{\theta},\underline{\theta}_0)}. \tag{3.15}$$

From Theorem 2.2, it is known that $Cov(u_{T,k}, u_{T,s})$ is approximately $\sigma^4$ (constant) for $k = s$ and approximately zero otherwise. Thus, treating the errors $u_{T,k}$, $k = 1, \cdots, \lfloor T/2 \rfloor$, as i.i.d., and in connection to (3.14), the right hand side of (3.15) tends to zero. This shows that (3.13) holds.

This completes the proof of Theorem 3.1. □

*Remarks.*

1. Note that the parameter space $\Theta \equiv \mathbf{R}_+^2 \times (0, 1/2)$ is not bounded. In this case, Theorem 2 in Wu (1981) cannot be applied.

2. Since $(a+b+c)^2 \leq 3(a^2 + b^2 + c^2)$, it can be seen from Lemma A.2 (part 2 and 5) that

$$k_T(\underline{\theta}, \underline{\theta}_0) \leq \lfloor T/2 \rfloor (\sigma_0^2 - \sigma^2)^2 + \frac{1}{2} T^2 (\mu_0^2 - \mu^2)^2 \left( \frac{2\lambda^3}{3} - \lambda^4 \right) + \frac{1}{2} T^2 (\lambda_0 - \lambda)^2 (\lambda + \lambda_0 - (\lambda + \lambda_0)^2).$$

In this case, $\overline{\lim}_{T \to \infty} T^{-2} k_T(\underline{\theta}, \underline{\theta}_0) = c$, $c > 0$. Therefore assumptions $A$ and $A'$ are not satisfied for the specific function $f(k, \underline{\theta}) = \sigma^2 + \mu^2 g(\lambda, k)$, in which case Theorem 3 in Wu (1981) cannot be applied. The same conclusion can be derived for Theorem 4 in Wu (1981).

3. If $\sigma_0^2 = \sigma^2$ i.e., there is no change in $\sigma^2$, then the condition in Theorem 3.1 is modified to

$$\frac{\mu_0^2}{\mu^2} \neq \frac{\lambda - \lambda^2}{\lambda_0 - \lambda_0^2}.$$

**Theorem 3.2.** *Assume that model (3.2) holds. If $E\|u\|^{2+\delta} < \infty$, for some $\delta > 0$, then for $\lambda \in (0, 1/2)$,*

$$\begin{pmatrix} \sqrt{T}(\hat{\sigma}^2 - \sigma^2) \\ T(\hat{\mu}^2 - \mu^2) \\ T(\hat{\lambda} - \lambda) \end{pmatrix} \to^L N(\underline{0}, \sigma^4 \Sigma),$$

where $\Sigma = \begin{pmatrix} 2 & 0 & 0 \\ 0 & 12/\lambda^3 & -6/\mu^2 \lambda^2 \\ 0 & -6/\mu^2 \lambda^2 & 2(2-3\lambda)/\mu^4 \lambda(1-2\lambda) \end{pmatrix}.$

**Proof.** Under the additional moment conditions of the error terms, then, following Fuller (pp.260-266, 2009), the asymptotic normality can be immediately seen. To obtain the variance-covariance matrix of $\hat{\underline{\theta}}_T$, we observe the following

$$\Sigma_T^{-1} = \begin{pmatrix} 1/\sqrt{T} & 0 & 0 \\ 0 & 1/T & 0 \\ 0 & 0 & 1/T \end{pmatrix} J'J \begin{pmatrix} 1/\sqrt{T} & 0 & 0 \\ 0 & 1/T & 0 \\ 0 & 0 & 1/T \end{pmatrix} \sim \begin{pmatrix} 1/2 & 0 & 0 \\ 0 & \lambda^3\left(\frac{1}{3} - \frac{\lambda}{2}\right) & \lambda^2\mu^2\left(\frac{1}{2} - \lambda\right) \\ 0 & \lambda^2\mu^2\left(\frac{1}{2} - \lambda\right) & \lambda\mu^4(1 - 2\lambda) \end{pmatrix} = \Sigma^{-1},$$

which leads to the desired results.

This completes the proof of Theorem 3.1. □

## 4. Simulation study

Here, the goal is to carry out a simulation study, mainly to empirically assess the performance of the spectral method for estimating the underlying parameters including the change-point. While doing so, it is also of interest to ascertain that the spectral method is robust to deviations from the Gaussian model. Thus, while performing the simulations we generate data from normal, $t_3$ and $\chi_1^2$ distributions. Our general observation has been that the method performs well for large values of $T$, the sample size. While we performed simulations for sample sizes $T = 2^8$, $2^9$, $2^{10}$, $2^{11}$, we present here results for only for $T = 2^{10}$. As for the change-point, we considered $\lambda = 0.20$, $0.25$, $0.30$, $0.35$, and $0.40$. The results for other sample sizes were not much different. The data generation process is based upon the model

$$X_t = \begin{cases} \nu_B + \xi_B Z_t, & t = 1, \cdots, \lfloor \lambda T \rfloor - 1 \\ \nu_A + \xi_A Z_t, & t = \lfloor \lambda T \rfloor, \cdots, T - 1, \end{cases} \tag{4.1}$$

where $Z_1, \cdots, Z_{T-1}$ are i.i.d. random variables following one of $N(0,1)$, $t_3$, or $\chi_1^2$ distributions. For each of the three distributions, values for the parameters $(\nu_B, \xi_B, \nu_A, \xi_A)$ are chosen in such a way that the choices ultimately lead to $\mu_B = 0$; $\sigma_B = 1$ and $\mu_A = 1.5$, $2.0$, and $2.5$; $\sigma_A = 1.2$, $1.6$, and $2.0$. Then, for each combination of $(\mu_B, \sigma_B, \mu_A, \sigma_A, \lambda)$, we computed values of $\underline{\theta} = (\mu^2, \sigma^2, \lambda)'$ and estimated the corresponding values through equations (3.5)-(3.7) iteratively under each of $N(0,1)$, $t_3$ and $\chi_1^2$ distributions. The estimation is based on 1000 simulations and the estimated values are presented in Table 1.

It is clear from Table 1 that all parameter estimates are quite close to true values for all three distributions. The parameter estimates are both good and robust to deviations from normality. However, it is interesting to note that values of $\hat{\sigma}^2$ are closer to $\sigma^2$ relatively under $N(0,1)$, and $\chi_1^2$ distributions compared to $t_3$ distribution, particularly for larger values of $\lambda$.

4. **Table 1:** Parameter estimates based on 1000 simulations under each of $N(0,1)$, $t_3$, and $\chi_1^2$ distributions when sample size $T = 2^{10}$.

| True Parameters | | | Normal Distribution | | | $t_3$ - Distribution | | | $\chi_1^2$ - Distribution | | |
|---|---|---|---|---|---|---|---|---|---|---|---|
| $\mu^2$ | $\sigma^2$ | $\lambda$ | $\hat{\mu}^2$ | $\hat{\sigma}^2$ | $\hat{\lambda}$ | $\hat{\mu}^2$ | $\hat{\sigma}^2$ | $\hat{\lambda}$ | $\hat{\mu}^2$ | $\hat{\sigma}^2$ | $\hat{\lambda}$ |
| 2.25 | 1.3520 | 0.2000 | 2.2367 | 1.3530 | 0.2044 | 2.2312 | 1.3386 | 0.2039 | 2.2488 | 1.3578 | 0.2037 |
| | 1.3300 | 0.2500 | 2.2419 | 1.3276 | 0.2540 | 2.2461 | 1.3127 | 0.2522 | 2.2501 | 1.3309 | 0.2519 |
| | 1.3080 | 0.3000 | 2.2632 | 1.3088 | 0.3018 | 2.2399 | 1.3030 | 0.3028 | 2.2669 | 1.3051 | 0.3024 |
| | 1.2860 | 0.3500 | 2.2637 | 1.2865 | 0.3509 | 2.2629 | 1.3087 | 0.3500 | 2.2607 | 1.2787 | 0.3495 |
| | 1.2640 | 0.4000 | 2.2562 | 1.2639 | 0.3989 | 2.2571 | 1.2325 | 0.4003 | 2.2444 | 1.2623 | 0.4006 |
| | 2.2480 | 0.2000 | 2.2552 | 2.2501 | 0.2056 | 2.2293 | 2.2403 | 0.2073 | 2.2575 | 2.2565 | 0.2034 |
| | 2.1700 | 0.2500 | 2.2437 | 2.1671 | 0.2552 | 2.2504 | 2.2075 | 0.2542 | 2.2760 | 2.1633 | 0.2529 |
| | 2.0920 | 0.3000 | 2.2540 | 2.0913 | 0.3036 | 2.2769 | 2.0970 | 0.3024 | 2.2616 | 2.0907 | 0.3033 |
| | 2.0140 | 0.3500 | 2.2691 | 2.0102 | 0.3491 | 2.2753 | 2.0165 | 0.3500 | 2.2787 | 2.0100 | 0.3493 |
| | 1.9360 | 0.4000 | 2.2571 | 1.9390 | 0.3995 | 2.2707 | 1.9173 | 0.4001 | 2.2570 | 1.9358 | 0.4002 |
| | 3.4000 | 0.2000 | 2.2573 | 3.3962 | 0.2083 | 2.2655 | 3.3267 | 0.2086 | 2.2588 | 3.4035 | 0.2092 |
| | 3.2500 | 0.2500 | 2.2703 | 3.2544 | 0.2584 | 2.2639 | 3.2051 | 0.2584 | 2.2684 | 3.2458 | 0.2561 |
| | 3.1000 | 0.3000 | 2.2742 | 3.0954 | 0.3041 | 2.2703 | 3.0171 | 0.3046 | 2.2464 | 3.0992 | 0.3060 |
| | 2.9500 | 0.3500 | 2.2717 | 2.9444 | 0.3504 | 2.2960 | 2.9354 | 0.3486 | 2.2674 | 2.9263 | 0.3482 |
| | 2.8000 | 0.4000 | 2.2656 | 2.8045 | 0.3996 | 2.2654 | 2.7692 | 0.3994 | 2.2602 | 2.8024 | 0.3990 |
| 4.00 | 1.3520 | 0.2000 | 4.0236 | 1.3533 | 0.2012 | 3.9825 | 1.3452 | 0.2023 | 3.9888 | 1.3518 | 0.2021 |
| | 1.3300 | 0.2500 | 3.9885 | 1.3297 | 0.2513 | 3.9814 | 1.3205 | 0.2516 | 4.0064 | 1.3226 | 0.2508 |
| | 1.3080 | 0.3000 | 4.0008 | 1.3059 | 0.3025 | 4.0022 | 1.3825 | 0.3022 | 4.0163 | 1.3122 | 0.3022 |
| | 1.2860 | 0.3500 | 4.0074 | 1.2846 | 0.3501 | 4.0039 | 1.2877 | 0.3501 | 4.0059 | 1.2824 | 0.3509 |
| | 1.2640 | 0.4000 | 4.0233 | 1.2616 | 0.4001 | 4.0202 | 1.2671 | 0.3997 | 4.0123 | 1.2576 | 0.4004 |
| | 2.2480 | 0.2000 | 3.9992 | 2.2452 | 0.2030 | 3.9782 | 2.5282 | 0.2051 | 4.0097 | 2.2686 | 0.2031 |
| | 2.1700 | 0.2500 | 3.9936 | 2.1680 | 0.2539 | 4.0049 | 2.1385 | 0.2517 | 4.0097 | 2.1852 | 0.2533 |
| | 2.0920 | 0.3000 | 3.9981 | 2.0908 | 0.3033 | 3.9932 | 2.0736 | 0.3024 | 4.0071 | 2.0854 | 0.3027 |
| | 2.0140 | 0.3500 | 4.0183 | 2.0132 | 0.3502 | 4.0284 | 1.9642 | 0.3495 | 4.0355 | 2.0154 | 0.3502 |
| | 1.9360 | 0.4000 | 4.0004 | 1.9372 | 0.3998 | 4.0196 | 1.8945 | 0.4005 | 4.0085 | 1.9426 | 0.4002 |
| | 3.4000 | 0.2000 | 3.9787 | 3.4003 | 0.2063 | 3.9975 | 3.3518 | 0.2059 | 3.9958 | 3.4082 | 0.2062 |
| | 3.2500 | 0.2500 | 3.9890 | 3.2483 | 0.2563 | 3.9941 | 3.2181 | 0.2537 | 3.9765 | 3.2368 | 0.2546 |
| | 3.1000 | 0.3000 | 4.0133 | 3.0971 | 0.3020 | 4.0476 | 3.0228 | 0.3003 | 4.0242 | 3.1102 | 0.3021 |
| | 2.9500 | 0.3500 | 4.0051 | 2.9534 | 0.3501 | 4.0473 | 2.8911 | 0.3491 | 4.0009 | 2.9522 | 0.3509 |
| | 2.8000 | 0.4000 | 4.0044 | 2.8018 | 0.3993 | 4.0165 | 3.0296 | 0.4007 | 4.0194 | 2.8124 | 0.4006 |
| 6.25 | 1.3520 | 0.2000 | 6.2460 | 1.3530 | 0.2005 | 6.2430 | 1.3261 | 0.2010 | 1.3504 | 6.2581 | 0.2005 |
| | 1.3300 | 0.2500 | 6.2674 | 1.3275 | 0.2508 | 6.2285 | 1.3180 | 0.2510 | 1.3304 | 6.2181 | 0.2511 |
| | 1.3080 | 0.3000 | 6.2511 | 1.3068 | 0.3019 | 6.2517 | 1.2903 | 0.3010 | 1.3087 | 6.2418 | 0.3027 |
| | 1.2860 | 0.3500 | 6.2696 | 1.2859 | 0.3504 | 6.2568 | 1.2721 | 0.3507 | 1.2807 | 6.2577 | 0.3504 |
| | 1.2640 | 0.4000 | 6.2436 | 1.2636 | 0.4004 | 6.2531 | 1.3139 | 0.4003 | 1.2689 | 6.2369 | 0.4006 |
| | 2.2480 | 0.2000 | 6.2360 | 2.2450 | 0.2014 | 6.2306 | 2.2402 | 0.2022 | 6.2498 | 2.2481 | 0.2015 |
| | 2.1700 | 0.2500 | 6.2441 | 2.1713 | 0.2508 | 6.2081 | 2.1699 | 0.2514 | 6.2622 | 2.1756 | 0.2503 |
| | 2.0920 | 0.3000 | 6.2628 | 2.0875 | 0.3013 | 6.2536 | 2.0962 | 0.3021 | 6.2522 | 2.0997 | 0.3036 |
| | 2.0140 | 0.3500 | 6.2682 | 2.0129 | 0.3504 | 6.2672 | 1.9831 | 0.3504 | 6.2641 | 2.0139 | 0.3510 |
| | 1.9360 | 0.4000 | 6.2416 | 1.9371 | 0.4007 | 6.2514 | 1.9180 | 0.4005 | 6.2617 | 1.9294 | 0.4001 |
| | 3.4000 | 0.2000 | 6.2083 | 3.4029 | 0.2040 | 6.2468 | 3.3308 | 0.2028 | 6.2071 | 3.3952 | 0.2031 |
| | 3.2500 | 0.2500 | 6.2462 | 3.2494 | 0.2522 | 6.2351 | 3.2087 | 0.2537 | 6.2672 | 3.2820 | 0.2522 |
| | 3.1000 | 0.3000 | 6.2639 | 3.0917 | 0.3025 | 6.2422 | 3.0114 | 0.3011 | 6.2837 | 3.0843 | 0.3030 |
| | 2.9500 | 0.3500 | 6.2618 | 2.9458 | 0.3496 | 6.2432 | 2.8577 | 0.3508 | 6.2700 | 2.9455 | 0.3499 |
| | 2.8000 | 0.4000 | 6.2584 | 2.8047 | 0.4006 | 6.2400 | 2.7645 | 0.4008 | 6.2565 | 2.7931 | 0.4008 |

## 5. Example: Well-log data

The well-log data was first considered by Ó Ruanaidh and Fitzgerald (1996). The data consists of 4050 measurements of the nuclear magnetic response of underground rocks when a probe is lowered into a bore-hole in the earth's surface. The data resembles a piecewise constant phenomenon with each segment relating to a single type of rock. So jump discontinuities occur in the data whenever the probe comes across a new type of rock. The determination of locations where the rock formation changes is important in the search for oil reserves. This is mainly to prevent blowouts, which may occur in the form of sudden and uncontrolled flows of drilling fluid, oil or water, up the borehole. Such blowouts can be avoided by adjusting the pressure in the borehole whenever a new type of rock is met. The detection of changes in rock strata as drilling proceeds is an important problem in oil discovery.

Here, we consider only the first 1500 data points as we search for change-points in the well-log data. The data (1500 data points) as collected are presented in Figure 1. Clearly, one must deal with the presence of outliers prior to proceeding with any analysis. For this purpose we proceed to identify outliers applying a filter proposed by Fearnhead and Clifford (2003), and also adopted by Wyse et al. (2011). Overall, 37 outliers have been identified through the filtering process and these outliers are replaced by simulating data points via normal distribution with mean and standard deviation determined by five observations before and after the outliers. This 'clean' data is presented in Figure 2.

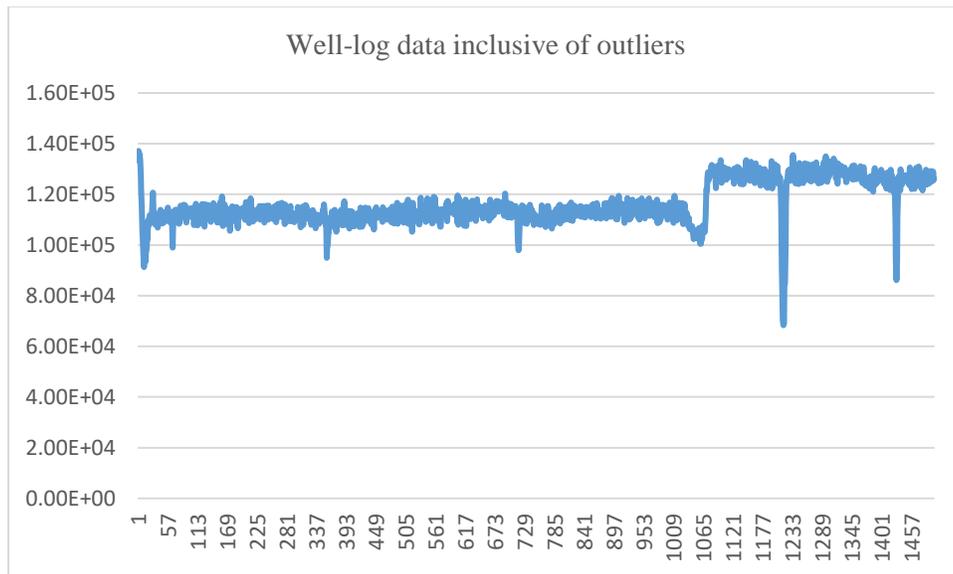

**Figure 1:** Well-log data (first 1500 observations) inclusive of all outliers

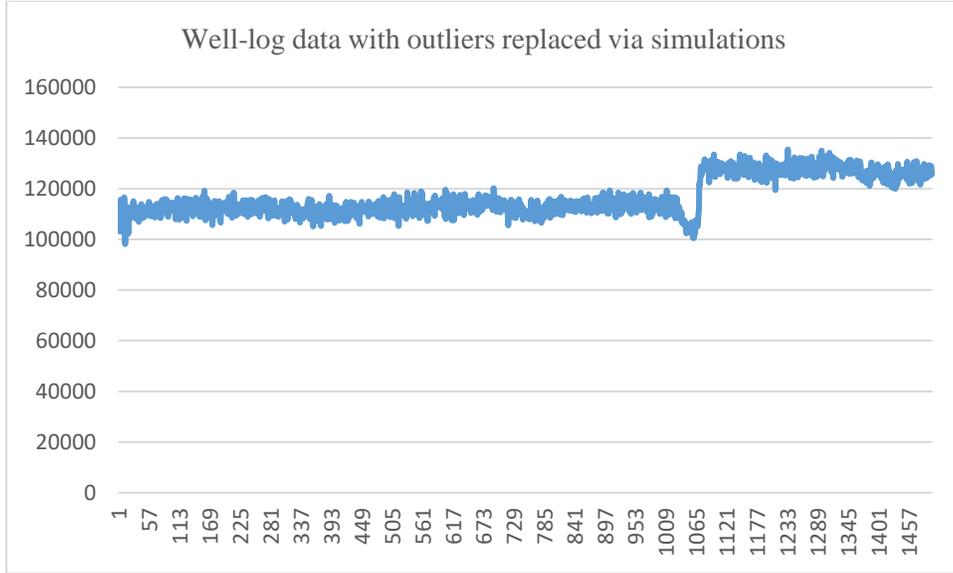

**Figure 2:** Well-log data with outliers replaced by simulations (using five observations before and after)

Likelihood ratio change detection methods proposed by Csörgo and Horváth (1997) show strong presence of a change-point (details omitted). The goal here is one of estimating the unknown change-point for the data in both Figures 1 and 2. For this purpose, let the data in either of the figures be represented by $\underline{X}_B = (X_0, X_1, \cdots, X_{\tau-1})$ with $T = 1501$. Let $\tau$ represent the change-point such that $X_0, X_1, \cdots, X_{\tau-1}$ have mean and variance given by $(\mu_B, \sigma_B^2)$ and $\underline{X}_A = (X_\tau, X_{\tau+1}, \cdots, X_{T-1})$ have mean and variance $(\mu_A, \sigma_A^2)$. We then pursue change-point estimation by considering the change-point as a proportion $\lambda_T = \tau/T$ such that $\lambda_T \in (0,1)$. Following the spectral analysis developed in the article, we let $\mu^2 = (\mu_B - \mu_A)^2$, and $\sigma^2 = \lambda \sigma_B^2 + (1-\lambda)\sigma_A^2$. The goal is to estimate the parameters $\underline{\theta} = (\sigma^2, \mu^2, \lambda)'$. As indicated earlier, one may note the change-point parameter $\lambda$ is first restricted so that $\lambda \in (0, 1/2)$. Upon obtaining $\hat{\lambda}$, one computes $\hat{\sigma}^2(\hat{\lambda}) = \hat{\lambda}\hat{\sigma}_B^2(\hat{\lambda}) + (1-\hat{\lambda})\hat{\sigma}_A^2(\hat{\lambda})$ and $\hat{\sigma}^2(1-\hat{\lambda}) = (1-\hat{\lambda})\hat{\sigma}_B^2(1-\hat{\lambda}) + \hat{\lambda}\hat{\sigma}_A^2(1-\hat{\lambda})$, and then one determines the proper estimate of the change-point as $\hat{\lambda}$ if $\hat{\sigma}^2(\hat{\lambda}) < \hat{\sigma}^2(1-\hat{\lambda})$. Otherwise, the parameter estimate is $1-\hat{\lambda}$. Estimates of all parameters $\underline{\theta} = (\sigma^2, \mu^2, \lambda)'$ for data in both Figure 1 and Figure 2 are presented in Table 2.

Table 2: Parameter estimates under spectral method for both original data in Figure 1 and clean data in Figure 2

| Data | $\hat{\sigma}^2$ | $\hat{\mu}^2$ | $\hat{\lambda}$ |
|---|---|---|---|
| **Original** | 9060160.0 | 187038300.0 | 0.7142 |
| **Clean** | 10250130.0 | 226933600.0 | 0.7242 |

It can be seen from Table 2 that change-point estimates for both original data and clean data are nearly the same. Thus, spectral method of change-point estimation is quite robust even under the presence of a large number of outliers. Such robustness was also seen in the simulations performed in Section 4.

**Appendix**

To evaluate the components in the matrix (3.11), the following two lemmas are necessary. The first lemma will be given without proof.

**Lemma A.1.** *For $k = 0$ we have*

1. $\lim_{k \to 0} \dfrac{\sin^2(\pi k \lambda)}{\sin^2(\pi k/T)} = \lambda^2 T^2$,

2. $\lim_{k \to 0} \dfrac{k \sin(2\pi k \lambda)}{\sin^2(\pi k/T)} = \dfrac{2\lambda T^2}{\pi}$,

3. $\lim_{k \to 0} \dfrac{\sin^4(\pi k \lambda)}{\sin^4(\pi k/T)} = \lambda^4 T^4$

4. $\lim_{k \to 0} \dfrac{k \sin(2\pi k \lambda) \sin^2(\pi k \lambda)}{\sin^4(\pi k/T)} = \dfrac{2\lambda^3 T^4}{\pi}$,

5. $\lim_{k \to 0} \dfrac{k^2 \sin^2(2\pi k \lambda)}{\sin^4(\pi k/T)} = \dfrac{4\lambda^2 T^4}{\pi^2}$.

**Lemma A.2.** *For $\lambda \in (0, 1/2)$, the following asymptotics are true*

1. $\langle 1, \underline{g}(\lambda, \underline{\omega}_T) \rangle = T^{-1} \sum_{k=1}^{T/2} \dfrac{\sin^2(\lambda \pi k)}{\sin^2(\pi k/T)} = \dfrac{\lambda - \lambda^2}{2} T + O(1)$,

2. $\langle \underline{g}(\lambda, \underline{\omega}_T), \underline{g}(\lambda, \underline{\omega}_T) \rangle = T^{-2} \sum_{k=1}^{T/2} \dfrac{\sin^4(\lambda \pi k)}{\sin^4(\pi k/T)} = \dfrac{1}{2}\left(\dfrac{2\lambda^3}{3} - \lambda^4\right) T^2 + O(T)$,

3. $\left\langle 1, \dfrac{\partial \underline{g}(\lambda,\underline{\omega}_T)}{\partial \lambda} \right\rangle = T^{-1} \sum_{k=1}^{T/2} \dfrac{k\pi \sin(2\lambda\pi k)}{\sin^2(\pi k/T)} = \left(\dfrac{1}{2} - \lambda\right) T + O(1),$

4. $\left\langle \underline{g}(\lambda,\underline{\omega}_T), \dfrac{\partial \underline{g}(\lambda,\underline{\omega}_T)}{\partial \lambda} \right\rangle = T^{-2} \sum_{k=1}^{T/2} \dfrac{k\pi \sin(2\lambda\pi k) \sin^2(\lambda\pi k)}{\sin^4(\pi k/T)} = \dfrac{1}{2}\left(\lambda^2 - 2\lambda^3\right) T^2 + O(T),$

5. $\left\langle \dfrac{\partial \underline{g}(\lambda,\underline{\omega}_T)}{\partial \lambda}, \dfrac{\partial \underline{g}(\lambda,\underline{\omega}_T)}{\partial \lambda} \right\rangle = T^{-2} \sum_{k=1}^{T/2} \dfrac{k^2 \pi^2 \sin^2(2\lambda\pi k)}{\sin^4(\pi k/T)} = \left(\lambda - 2\lambda^2\right) T^2 + O(T).$

**Proof**. To obtain the approximations in Lemma 3.2, we consider summations from 0 to $T-1$. Next, to obtain parts (1) through (5) the symmetry is applied, i.e., we split the sums to half and exclude the zero term.

Note that $\left|\left(1 - e^{i\tau\omega_j^{(T)}}\right)\Big/\left(1 - e^{i\omega_j^{(T)}}\right)\right| = \left(1 - \cos(\tau\omega_j^{(T)})\right)\Big/\left(1 - \cos(\omega_j^{(T)})\right) = \sin^2(\tau\omega_j^{(T)}/2)\Big/\sin^2(\omega_j^{(T)}/2).$

Letting $\omega_j^{(T)} = 2\pi j/T$, $j = 0, 1, \cdots, T-1$, it can also be seen that

$$\dfrac{\partial}{\partial \lambda}\left\{\left|\left(1 - e^{i\lambda 2\pi k}\right)\Big/\left(1 - e^{i2\pi k/T}\right)\right|\right\} = 2\pi k \sin(\lambda 2\pi k)\Big/\sin^2(\pi k/T).$$

Hence, it follows that

$$T^{-1} \sum_{k=0}^{T-1} \left|\dfrac{1 - e^{i\lambda 2\pi k}}{1 - e^{i2\pi k/T}}\right|^2 = T^{-1} \sum_{k=0}^{T-1} \left|\sum_{m=0}^{\tau-1} e^{im\omega_k^{(T)}}\right|^2 = T^{-1} \sum_{k=0}^{T-1} \sum_{m,n=0}^{\tau-1} e^{i(m-n)\omega_k^{(T)}}$$

$$= T^{-1} \sum_{k=0}^{T-1} \left\{\tau + \sum_{\substack{m,n=0 \\ m\neq n}}^{\tau-1} e^{i(m-n)\omega_k^{(T)}}\right\} = T^{-1}\left\{T\tau + \sum_{\substack{m,n=0 \\ m\neq n}}^{\tau-1} \sum_{k=0}^{T-1} e^{i(m-n)\omega_k^{(T)}}\right\}$$

$$= \lambda_T T.$$

(A.1.1)

Also, note that

$$T^{-2} \sum_{k=0}^{T-1} \left|\dfrac{1 - e^{i\lambda 2\pi k}}{1 - e^{i2\pi k/T}}\right|^4 = T^{-2} \sum_{k=0}^{T-1} \left|\sum_{m=0}^{\tau-1} e^{im\omega_k^{(T)}}\right|^4 = T^{-2} \sum_{k=0}^{T-1} \left\{\tau + \sum_{\substack{m,n=0 \\ m\neq n}}^{\tau-1} e^{i(m-n)\omega_k^{(T)}}\right\}^2$$

$$= T^{-2} \sum_{k=0}^{T-1} \left\{\tau^2 + 2\tau \sum_{\substack{m,n=0 \\ m\neq n}}^{\tau-1} e^{i(m-n)\omega_k^{(T)}} + \left(\sum_{\substack{m,n=0 \\ m\neq n}}^{\tau-1} e^{i(m-n)\omega_k^{(T)}}\right)^2\right\}$$

$$= T^{-2}\tau^2 T + T^{-2} \sum_{k=0}^{T-1} \left(\sum_{\substack{m,n=0 \\ m\neq n}}^{\tau-1} e^{i(m-n)\omega_k^{(T)}}\right)^2 = T^{-1}\tau^2 + 2T^{-2} \sum_{k=0}^{T-1} \sum_{j=1}^{\tau-1}(\tau - j)^2$$

$$= \dfrac{\tau^2}{T} + \dfrac{(\tau-1)\tau(2\tau-1)}{3T} = T^2 \dfrac{2\lambda_T^3}{3} + \dfrac{\lambda_T}{3}.$$  (A.1.2)

In A.2, the following identity was utilized

$$\left(\sum_{\substack{m,n=0\\m\neq n}}^{\tau-1} e^{i(m-n)\omega_k^{(T)}}\right)^2 = \left(\sum_{\substack{m,n=0\\m<n}}^{\tau-1} e^{i(m-n)\omega_k^{(T)}} + \sum_{\substack{m,n=0\\m>n}}^{\tau-1} e^{i(m-n)\omega_k^{(T)}}\right)^2$$

$$= \left\{\sum_{j=1}^{\tau-1}(\tau-j)e^{ij\omega_k^{(T)}} + \sum_{j=1}^{\tau-1}(\tau-j)e^{-ij\omega_k^{(T)}}\right\}^2 = \left\{\sum_{j=1}^{\tau-1}(\tau-j)\left(e^{ij\omega_k^{(T)}} + e^{-ij\omega_k^{(T)}}\right)\right\}^2$$

$$= \sum_{j=1}^{\tau-1}(\tau-j)^2\left(2 + 2\operatorname{Re} e^{i2j\omega_k^{(T)}}\right) + 2\operatorname{Re}\sum_{\substack{m,n=1\\m>n}}^{\tau-1}(\tau-m)(\tau-n)\left(e^{i(m-n)\omega_k^{(T)}} + e^{i(m+n)\omega_k^{(T)}}\right).$$

Also note that the real part of a complex number $z$ is $(z+\bar{z})/2$, thus it is enough to show that

$$\sum_{k=0}^{T-1}\sum_{j=1}^{\tau-1}(\tau-j)^2 e^{i2j\omega_k^{(T)}} = \sum_{j=1}^{\tau-1}(\tau-j)^2 \sum_{k=0}^{T-1} e^{i2j\omega_k^{(T)}} = 0,$$

That the expression above is zero follows from the orthonormality property.

Applying a derivative with respect to $\lambda_T$ into (A.1.1) and (A.1.2), in combination with Lemma A.1, the resulting summations will assist us to obtain parts 3 through 5.

This completes the proof of Lemma A2. □